\documentstyle[12pt,twoside,fleqn,espcrc1]{article}
\newcommand{\AmS}{{\protect\the\textfont2
  A\kern-.1667em\lower.5ex\hbox{M}\kern-.125emS}}

\begin{document}
\title {\null\vskip -1.0cm\hfill {\small ORNL-CTP-9703 and
hep-ph/9712320}
\\  \vskip 0.8cm Heavy Quarkonium Production and Propagation in Nuclei}

\author{Cheuk-Yin Wong\address{Physics Division, Oak Ridge National
Laboratory, Oak Ridge, TN 37831 }%
\thanks{ This research was supported by the
Division of Nuclear Physics, U.S. D.O.E.  under Contract
DE-AC05-96OR22464 managed by Lockheed Martin Energy Research Corp.}
}


\maketitle

\begin{abstract}

We describe a precursor in heavy quarkonium production in terms of a
coherent admixture of states of different color, spin, and angular
momentum quantum numbers, and obtain the production amplitudes for
different quarkonium bound states by projecting out this precursor
state onto these bound states.  The precursor is absorbed in its
passage through a nucleus in a $pA$ reaction, and the total cross
section between this precursor with a nucleon can be calculated with
the two-gluon model of the Pomeron.  Such a description of coherent
precursors and their subsequent interactions with nucleons can explain
many salient features of $J/\psi$ and $\psi'$ production in $pA$
collisions.

\end{abstract}

\section{ Introduction } 

It is a great pleasure to write this article in honor of Prof. Tai-You
Wu's 90th Birthday.  Prof.\ Wu's grandfather and father were
respectively ``Jin Shi'' and ``Ju Ren'', which were among the highest
honors a Chinese scholar could achieve in the last century.\break
Prof. Wu himself was instrumental in bringing modern physics to China,
and has trained a large number of physicists, including C. N. Yang and
T. D. Lee.  Through the lineage of Tai-You Wu, Chinese physics
scholars of today are a part of the continuation of the Chinese
scholarship of past few thousand years, and the solid record of
accomplishments of past Chinese scholars will inspire Chinese
physicists to continue to make contributions to the world of science
in the future.

In the search of for the quark-gluon plasma, it has been suggested
that the production of charmonium will be suppressed in a quark-gluon
plasma because of the screening of the interaction between $c$ and
$\bar c$ \cite{Mat86}.  To extract information on the suppression due
to the quark-gluon plasma, it is necessary to study the suppression of
$J/\psi$ production by sources different from the quark-gluon plasma.
It is therefore useful to examine the mechanism of heavy quarkonium
production and its propagation in nuclei
\cite{Ger88,Won96,Won96a,Kha96,Kha96a,Gav90,Huf96,Woncw96b,Woncw96a,Qia97}.

To follow the processes of production and propagation of a charmonium
or a heavy quarkonium, one needs to understand the nature of the
object that is formed by the collision of partons in a hard-scattering
process.  This leads to the picture of a precursor formed by parton
collisions that is a coherent mixture of different color, spin, and
angular momentum states.  The production and the subsequent
interaction of this object with nucleons along its passage through a
nucleus is the subject of the present study.

The present work is the result of a continuing evolution and synthesis
of many previous attempts to understand production and the absorption
of heavy quarkonium in a nucleus.  Previous studies on heavy
quarkonium production include the color evaporation model
\cite{Fri77}, the color-singlet model \cite{Cha80}, and the
color-octet model \cite{Bod95}.  The absorption of $J/\psi$ has been
previously described as the propagation of a hybrid object $[(c\bar
c)_8 g]$ \cite{Kha96}, the additive quark model \cite{Won96}, a
color-singlet coherent state due to the multiple collisions with
nucleons \cite{Huf96,Woncw96b}, and an incoherent admixture of color
singlet and color-octet states \cite{Woncw96a,Qia97}.  The model we
shall present in terms of a coherent color admixture of the precursor
provides a description of the important features of the production and
absorption process.  There are also specific predictions of this model
which can be further tested by experiment.

\section{ Production Process } 

We can study the collision of the parton $b$ of a beam nucleon with
the parton $a$ of the target nucleon leading to the production of a
$Q$-$\bar Q$ pair which later materializes directly or indirectly into
a bound heavy quarkonium.  The initial state $\Phi_{a b}$ of the
$Q$-$\bar Q$ pair from the collision at $t_i=0$ is represented by the
state vector
\begin{eqnarray}
|\Phi_{a b}(t_i) >=
{\cal M}(ab \rightarrow Q(P/2+q){\bar Q}(P/2-q))
 |Q(P/2+q){\bar Q}(P/2-q)>
\end{eqnarray}
where ${\cal M}(ab \rightarrow Q(P/2+q){\bar Q}(P/2-q))$ is the
Feynman amplitude for the $a+b \rightarrow Q + \bar Q$ process, $P$ is
the center-of-mass momentum, $q$ is the relative momentum of the
$Q$-$\bar Q$ pair.  For simplicity of notation, the color and
azimuthal spin components are understood.

One can perform a decomposition in terms of color and angular momentum
states as
\begin{eqnarray}
\label{eq:adm}
|\Phi_{a b}(t_i) > 
=\sum_{C J L S }
{\tilde \phi}^C_{JLS}( q)   |Q {\bar Q}[{}^SL_J^C](P)> 
\end{eqnarray}
where 
\begin{eqnarray}
{\tilde \phi}^C_{JLS}({ q}) 
= <Q {\bar Q}[{}^SL_J^C](P)|{\cal M}(ab \rightarrow Q(P/2+q){\bar
Q}(P/2-q))|Q(P/2+q){\bar Q}(P/2-q)> 
\end{eqnarray}
with $|Q {\bar Q}[{}^SL_J^C](P)>$ describing the center-of-mass motion
of the $Q\bar Q$ pair in color state $C$, and angular momentum
quantum numbers $JLS$.  

The state $\Phi_{a b}$ will evolve according to perturbative QCD,
\begin{eqnarray}
\label{eq:evol}
|\Phi_{a b}(t) >~ = ~U(t,t_i)|\Phi_{a b}(t_i)> ~=~ T~\exp\{ -i
\int_{t_i}^t H_I dt\}~ |\Phi_{a b}(t_i) >
\end{eqnarray}
where the evolution operator $U(t,t_i)$ can be expanded out in a
time-ordered perturbation series in terms of the interaction $H_I$ .
The state $\Phi_{a b}$ can be called the heavy quarkonium precursor
state formed by partons $a$ and $b$.  Because the partons are
propagators whose legs joined onto their parent nucleons, the partons
are off the mass shell, with energies that depend on the invariant
masses of the other products of the nucleon-nucleon collision (see
\cite{Sch77} and Eq. (4.14) of \cite{Won94}). Thus, the energy of the
state $\Phi_{a b}$ also depends on the invariant masses of the other
products of the nucleon-nucleon collision.  It is reasonable to
associate the on-shell energies of the state $\Phi_{a b}$ as an
average energy of an admixture of bound states of different energies
and obtain the bound state production amplitude by projecting the
asymptotic precursor state at $t=\infty$ onto various bound states
within a certain energy range.  For example, guided by the evaporation
model, one can consider this range as between $2m_c$ and $2m_D$ for
$c\bar c$ quarkonium production.

We describe a bound $Q\bar Q$ state with the quantum numbers $JLS$ and
other quantum numbers with a center-of-mass momentum $P$ as
\begin{eqnarray}
\label{eq:c1}
|\Psi_{JLS};Pq>= \sqrt{2M_{JLS} \over 4m_Q m_{\bar Q}}
{\tilde R}_{JLS}({ q}) |Q {\bar Q}[{}^SL_J^{(1)}](P)>.
\end{eqnarray}
The above state may be specified by additional quantum numbers such as
the number of radial nodes, etc.  For simplicity of notation, these
additional quantum numbers will not be written out explicitly.

In the lowest-order perturbative QCD, the probability amplitude for
the direct production of $\Psi_{JLS}$ is obtained from projecting
$\Phi_{ab} (t_i) $ onto $\Psi_{JLS}$.  The projection is simplest in
the $Q$-$\bar Q$ center-of-mass system where $P=(M_{JLS},{\bf 0})$ and
$q=(0,{\bf q})$ and the probability amplitude is \cite{Pes95,Cra91}
\vskip -0.4cm
\begin{eqnarray}
\label{eq:overlap}
<\Psi_{JLS};Pq|\Phi_{a b}(t=\infty) >_{\rm ~lowest~order} &=&
<\Psi_{JLS};Pq|\Phi_{a b}(t_i) > \nonumber\\
&=& \sqrt{ 2M_{JLS} \over 4m_Q m_{\bar
            Q}} \int {d^3{\bf q} \over (2\pi)^3} {\tilde R}_{JLS}({\bf
            q}) {\tilde \phi}_{JLS}^{(1)}({\bf q}).
\end{eqnarray}
\vskip -.2cm
\noindent
where $M_{JLS}$ is the mass of the bound state, $m_Q$ the mass of the
quark, and we have followed the normalization of \cite{Pes95}.
Because the bound state $\Psi_{JLS}$ is a color-singlet state, the
above projection will involve only color-singlet components of the
admixture in Eq. (\ref{eq:adm}).

In the next-order perturbation theory, the color-singlet bound state
$\Psi_{JLS}$ accompanied by a gluon $g$ can be produced by the
color-octet component of $ \Phi_{ab}$ in Eq.\ ({\ref{eq:adm}).  The
probability amplitude for the production of the bound state
$\Psi(JLS)$ accompanied by a gluon is
\begin{eqnarray}
<[\Psi_{JLS};Pq]g|
\Phi_{ab}(t)>=<[\Psi_{JLS};Pq]g|U(t,t_i)| \Phi_{ab}(t_i)>.
\end{eqnarray}

The state $\Psi_{JLS}$ can also be produced indirectly through the
production of different bound states $\Psi_{J'L'S'}$ which
subsequently decay into $\Psi_{JLS}$.  For example, in $J/\psi$
production, a large fraction of the observed $J/\psi$ comes from the
radiative decay of $\chi_1$ and $\chi_2$ \cite{Ant93}.  The production
probability, including direct, indirect, and color octet contributions,
is then the sum of the absolute squares of various amplitudes.

Heavy quarkonia can be produced by different parton combinations such
as $g$-$g$,\break $q$-$\bar q$, and $g$-$q$ collisions, which will
lead to different precursor states.  The total production probability
will be the sum from all different precursor states.

For lack of a better name, we shall call the present model the
coherent precursor model.  Such a model has many features similar to
the color evaporation model.  In fact, one can interpret the
phenomenological fractional coefficient in the color evaporation model
as the absolute square of various projection amplitudes.  The
approximate success of the color evaporation model \cite{Kow94} also
assures that this projection treatment will lead to a good description
of many pieces of experimental data.

\section{Interaction of a coherent precursor with a nucleon} 

The interaction of the precursor with hadron matter or other medium
depends on the relative kinetic energy of this precursor with respect
to the hadron matter or the medium.  For processes when the precursor
travels at a high energy relative to hadron matter, the interaction of
the precursor with respect to a hadron in the medium can be described
in terms of Pomeron and Reggeon exchanges. On the other hand, at
energies where the precursor is nearly at rest with respect to the
medium, the interaction depends on the binding energies of various
bound states.

We shall consider here the case of charmonium production in $pA$
collisions at a fixed target energy of a few hundred GeV
\cite{Bad83,Ald91}, and $\Upsilon$ production at 800 GeV
\cite{Ald91a}.  At these energies, the production of a heavy
quarkonium state with $x_F>0$ will involve the interaction of the
precursor with nuclear matter at high energies.  The total cross
section between the precursor and a model meson $M(q_3\bar q_4)$ can
be obtained by using the two-gluon model of the Pomeron
\cite{Low75,Nus75,Dol92,Woncw96b}.  Using the notation of
\cite{Dol92}, the elastic scattering amplitude for the exchange of two
gluons between the precursor $\Phi_{ab}(Q_1\bar Q_2)$ and the meson
$M(q_3\bar q_4)$ in this model is
\begin{eqnarray}
{\cal A}(s,t)={ig^4 s \over 16} \int d{\bf b} e^{-i{\bf Q}\cdot {\bf
b}}
<\Phi_{ab}(Q_1{\bar Q}_2) M(q_3 {\bar q}_4)|V|\Phi_{ab}(Q_1{\bar Q}_2)
M(q_3 {\bar q}_4)>
\end{eqnarray}
where $s=(p(\Phi_{ab},{\rm initial})+p(M(q_3\bar q_4),{\rm
initial}))^2$, $t=(p(\Phi_{ab},{\rm final})-p(\Phi_{ab},{\rm
initial}))^2=Q^2$.  For the exchange of two-gluons, $V$ is
\cite{Dol92}
\begin{eqnarray}
V=\sum_{a} [\lambda_1^a \lambda_3^a V_{13}
+ (\lambda_2^a)^T (\lambda_4^a)^T V_{24}
- \lambda_1^a (\lambda_4^a)^T V_{14}
- (\lambda_2^a)^T \lambda_3^a V_{13}]^2,
\end{eqnarray}
where $\lambda_i^a$ is the generator of SU(3) associated with the
$i$th quark, $V_{ij}$ is the interaction between the $i$ and the $j$th
quark or antiquark.  Because $(q_3\bar q_4)$ is coupled to a
color-singlet state, the matrix element in the above equation between
different color states vanishes, and we have
\begin{eqnarray}
\label{eq:ast}
&{}& {}\!\!\!\!\!\!\!{\cal A}(s,t) \\ &=& \!\!\!\!\! \sum_{CJLS}
\!\!\!  {ig^4 s \over 16} \! \int d{\bf b} d{\bf r}_{12} d{\bf r}_{34}
e^{-i{\bf Q}\cdot {\bf b}} |\phi_{JLS}^C ({\bf r}_{12})|^2 | R({\bf
r}_{34})|^2 \!\!\! < \!\! (Q_1\bar Q_2)^C (q_3 \bar q_4)^1|V| \! (
\!Q_1\bar Q_2)^C (q_3 \bar q_4)^{1} \!\!\!>, \nonumber
\end{eqnarray}
where $R({\bf r}_{34})$ is the $q_3$-$\bar q_4$ relative wave function
in the model meson $M(q_3\bar q_4)$.  One can use the optical theorem
to obtain the total cross section for this coherent precursor. One can
compare the total cross section $\sigma_{JLS}^C(ab)$ obtained from the
forward elastic scattering amplitude for the case if $\Phi_{ab}$ is a
(properly normalized) pure state of $\phi_{JLS}^C({\bf r}_{12}
)|(Q_1\bar Q_2[{}^S L_J^C] (P)>$. Then Eq.\ (\ref{eq:ast}) gives the
total cross section between the precursor (formed by the partons $a$
and $b$) given by
\begin{eqnarray}
\sigma_{tot}(ab)=\sum_{CJLS} f_{JLS}^C(ab) ~\sigma_{JLS}^C(ab) 
\end{eqnarray}
\vskip -0.5cm
\begin{eqnarray}
\label{eq:result}
{\rm where~~~~} 
f_{JLS}^C(ab) = \int d{\bf r}_{12} |\phi_{JLS}^C ({\bf r}_{12}) |^2 /<\Phi_{ab}
|\Phi_{ab}>.
\end{eqnarray}
\vskip -0.2cm
The above describes the cross section for a coherent precursor
colliding with a meson.  The result of Eq. (\ref{eq:result}) can also
be generalized to that for the collision between the precursor and a
nucleon as is done in \cite{Dol92}.

From the work of Dolej\v s\' i and H\"ufner \cite{Dol92}, and Wong
\cite{Woncw96b}, we know that the $(Q\bar Q)^{(1)}$-nucleon total
cross section for a pure color-singlet $(Q\bar Q)^{(1)}$ state is
approximately proportional to the root-mean-square of the separation
between $Q$ and $\bar Q$.  The coherent precursor produced after the
hard-scattering process has a small spatial extension, so it is
reasonable to take the color-singlet cross sections
$\sigma_{tot}^{(1)}$ for different $JLS$ states to be small and nearly
equal.  The $(Q\bar Q)^{(8)}$-nucleon total cross section for a pure
color-octet $(Q\bar Q)^{(8)}$ state is insensitive to the $Q$-$\bar Q$
separation.  This approximate independence of the color-octet cross
section with respect to the $Q$-$\bar Q$ separation \cite{Dol92,Won96}
allows us to approximate the color-octet cross sections for different
color-octet $JLS$ states by a single $\sigma_{tot}^{(8)}$.  Therefore,
the precursor-nucleon total cross section is
\begin{eqnarray}
\label{eq:resultp}
\sigma_{tot} (ab) \approx f^{(1)}(ab) ~ \sigma_{tot}^{(1)} +
f^{(8)}(ab)~ \sigma_{tot}^{(8)},
\end{eqnarray}
\begin{eqnarray}
\label{eq:f8a}
{\rm where~~~~}
f^C (ab) =\sum_{JLS} f_{JLS}^C (ab).
\end{eqnarray}

In heavy quarkonium production, there can be different combinations of
partons leading to different precursor states, and each precursor has
a total precursor-nucleon cross section given by an expression such as
Eq.\ (\ref{eq:resultp}).  Thus, the precursor cross section depends on
the color-octet fraction of the precursor.

The most important parton collisions for heavy quarkonium production
are the $gg$ and $q\bar q$ collisions.  In $gg$ collisions, the
coherent precursor can be in color-singlet and color-octet states,
while in $q \bar q$ collisions, the precursor is a color-octet state in
the lowest order.  For the total yield at fixed target energies, the
dominant production process comes from the collision of gluons
\cite{Kow94}.

In the kinematic region where the collision of the precursor with
nucleons takes place at high energies, as in the production of heavy
quarkonium in the region of $x_F>0$ in $pA$ collisions, the binding
energy of the heavy quarkonium is much smaller than the
precursor-nucleon collision energy.  The probability for the breakup
of the $Q$-$\bar Q$ pair into $Q\bar q$ and $q\bar Q$ by combining
with light quarks is large in a precursor-nucleon collision.  It is
reasonable to assume that in high-energy precursor-nucleon collisions
the absorption (i.e. breakup) cross section, $\sigma_{abs}(ab)$, is
approximately given by the precursor-nucleon total cross section
$\sigma_{tot} (ab)$ of Eq.\ (\ref{eq:resultp}).

\section{Propagation of a coherent precursor in nuclear matter} 

We consider a $pA$ collision in which a heavy quarkonium precursor is
produced in one of the nucleon-nucleon collisions and the precursor
subsequently travels in the nuclear medium of the target nucleus.
Because of the coherence of the precursor, it propagates through the
nuclear medium as a single object with a single absorption cross
section after its production.  When the precursor travels inside the
nuclear medium, the time of evolution $t$ can be represented
equivalently by the corresponding path length $L/v$, where $v$ is the
velocity of the precursor in the medium.  The state vector after
propagating a distance $L$ in the nuclear medium is related to the
state vector after production by
\begin{eqnarray}
|\Phi_{ab}(L) > = e^{-\rho \sigma_{abs}(ab) L/2}|\Phi_{ab}(L=0)>\,,
\end{eqnarray}
where $\sigma_{abs}(ab) \equiv \sigma_{abs}(\Phi_{ab}{\rm -}N)$ is the
precursor absorption cross section for the collision of the precursor
$\Phi_{ab}$ with a nucleon, and $\rho$ is the nuclear matter number
density.

After the passage through a path length of $L$ in the nuclear matter, the
amplitude for the production of a bound state $\Psi_{JLS}$ is obtained
by projecting the state vector $\Phi_{ab}(L)$ onto the bound state
$\Psi_{JLS}$.  Therefore,
\begin{eqnarray}
\label{eq:rat1}
<\Psi_{JLS};Pq|\Phi_{a b}(L) >
= e^{-\rho \sigma_{abs}(ab) L/2} <\Psi_{JLS};Pq|\Phi_{a b}(L=0) >
\end{eqnarray}
and for the octet production of $|\Psi_{JLS};Pq>$ with the emission of
a gluon
\begin{eqnarray}
\label{eq:rat2}
<[\Psi_{JLS};Pq]g|U(t,L/v)| \Phi_{ab}(L)>= e^{-\rho \sigma_{abs}(ab)
L/2}<[\Psi_{JLS};Pq]g|U(t,0)| \Phi_{ab}(L=0)>.
\end{eqnarray}
When we include precursors from different parton collisions leading to
the production of the bound state $\Psi_{JLS}$, there will be
different survival factors $e^{-\rho \sigma_{abs}(ab) L/2}$ for
different parton combinations $a$-$b$.  At fixed-target energies,
where the total yield of $J/\psi$ in the forward direction is
dominanted by contributions from $gg$ collisions \cite{Kow94}, one
expects that there is essentially only a single survival factor for
the total yield in forward directions.

\section{Consequences of the coherent precursor model}

One can outline a few distinct consequences of the model we have
presented.  In the situation when only a single absorption combination
of partons dominates the production process, the ratio of the
production of various bound states in a $pA$ collision will be
independent of the mass number of the nucleus.  This arises because
the precursor is absorbed in its passage through nuclear matter by a
single precursor-nucleon cross section $\sigma_{abs}(ab)$, and the
production of all different bound states comes from the projection of
the precursor state onto the bound states after the absorption.  For
example, as a consequence of Eqs.\ (\ref{eq:rat1}),
\begin{eqnarray}
{|<\psi'|\Phi_{gg}(L)>|^2 \over |<\psi|\Phi_{gg}(L)>|^2}
={|<\psi'|\Phi_{gg}(L=0)>|^2 \over |<\psi|\Phi_{gg}(L=0)>|^2}.
\end{eqnarray}
There is a similar relation for the production amplitude from the
color-octet amplitudes because of Eq.\ (\ref{eq:rat2}).  Therefore, we
have
\begin{eqnarray}
{\sigma(\psi') (L) \over \sigma(\psi) (L)}
=
{\sigma(\psi') (L=0) \over \sigma(\psi) (L=0)}
~~~~{\rm or ~~~~}
{\sigma(pA\rightarrow \psi'~X) \over \sigma(pA\rightarrow \psi~X) }
={\sigma(pp\rightarrow \psi'~X) \over \sigma(pp\rightarrow \psi~X) },
\end{eqnarray}
which is independent of $A$. A similar independence with $A$ is
predicted for $\Upsilon'/ \Upsilon$.  Experimentally, the ratio
$\psi'/(J/\psi)$ is observed to be approximately a constant of the
atomic numbers \cite{Ald91}, in agreement with the present picture.
The experimental yields of $\Upsilon$ and $\Upsilon'$ have
considerable uncertainties, but within the large experimental
uncertainties the ratio ${\sigma(pA\rightarrow \Upsilon'~X) /
\sigma(pA\rightarrow \Upsilon~X)}$ is approximately independent of $A$
\cite{Ald91a}.  The present model predicts further that the ratio
of the $\chi$ yield to the $J/\psi$ yield will also be independent of the mass
number in $pA$ collisions.  It will be of interest to test such a
prediction in the future.

A similar model of coherent state production was put forth earlier by
Dolej\v s\' i and H\"ufner \cite{Dol92}, and Wong \cite{Woncw96b},
with some major differences.  In these models, a color-singlet
coherent state of $J/\psi$ and $\psi'$ is assumed to come from the
successive multiple scattering of the precursor with a row of
nucleons, whereas in the present model, a coherent state of both
color, angular momentum, and other quantum numbers is formed in the
hard-scattering process before multiple collisions with nucleons and
collisions with nucleons at high energy are assumed to lead to
absorption, and not to the change of the coherent mixture.

We note that the total cross section $\sigma_{tot}$ for a coherent
color admixture is the weighted sum of the corresponding cross
sections for pure color-singlet and color-octet states.  For a
coherent state that is formed after the hard scattering, the spatial
extension of the state is small.  One expects that
$\sigma_{tot}^{(1)}$ should be nearly zero while $\sigma_{tot}^{(8)}$,
which is nearly independent of the radial size of the state, is very
large, of the order of 30 to 60 mb. Thus, an absorption cross section
of the order of 6 mb extracted from the experimental survival factor in
$pA$ data \cite{Ger88,Won96a,Kha96a} for $J/\psi$ and $\psi'$ can be
understood as arising from a color-octet fraction $f^{(8)}$ of the
order of 10\% to 20\%, as given by Eq.\ (\ref{eq:resultp}).

Experimentally, the production cross sections for $\chi_1$, $\chi_2$,
and direct $J/\psi$ are 131 nb, 188 nb, and 102 nb, respectively, for
$\pi$-N collisions at 300 GeV \cite{Ant93}.  Therefore, direct
$J/\psi$ production is about 25\% of the total yield of charmonium.
As $\chi_1$ and $\chi_2$ are produced predominantly via the
color-singlet states, while the direct $J/\psi$ production comes
predominantly from color-octet states \cite{Ben96,Tan96}, it is
therefore encouraging that the fraction of 25\% of direct $J/\psi$
among the total charmonium yield is of the same order as the
color-octet fraction $f^{(8)}$ of about 10\%-20\% that is consistent
with the absorption cross section and the theoretical estimates from
the two-gluon model of the Pomeron.  It will be of interest to
calculate this color-octet fraction $f^{(8)}$ from Eqs.\
(\ref{eq:result}) and (\ref{eq:f8a}) theoretically.  This may rectify
a previously unsatisfactory outcome when the precursor is described as
a mixed state with an incoherent color admixture.  The incoherent
color admixture description of the precursor is not satisfactory
because the absorption cross section for the color-octet state is
found, in this picture, to be about 15 to 20 mb in \cite{Woncw96a} and
11 mb in \cite{Qia97}, too small to be consistent with the theoretical
estimates of about 30-60 mb from the two-gluon model of the Pomeron
\cite{Dol92,Woncw96b}.

Finally, in different kinematic regions, the importance of the
different parton production components will change \cite{Kow94}.  For
example, while $gg$ fusion dominates in the region of small $x_F$,
which gives the largest yield, the yield at $x_F>>0.6$ is dominated by
$q$-$\bar q$ collisions.  Unlike the $gg$ fusion which can lead to
both color-singlet and color-octet components, $q\bar q$ collisions
lead to color-octet in the lowest-order approximation.  The present
model will predict an increase in the absorption cross section in the
region of large $x_F$ due to the large color-octet precursor-nucleon
cross section.  Experimental data at different $x_F$ show an increase
in the effective absorption cross section as $x_F$ increases
\cite{Woncw96a,Bad83,Ald91}.  However, the effect of energy loss needs
to be taken into account at these kinematic regions to confirm the
effect that the color-octet state leads to a very large increase in
the absorption of the precursor at $x_F>0.6$.

In the transition region when two different precursors contribute
about equally, as in the region around $x_f\sim 0.6$, the survival
probability is determined by two different exponential survival
factors, with two different absorption cross sections for the two
kinds of precursors, as a function of the nuclear matter path length.
Future experimental investigations on the change of the
characteristics of the mass dependence from small $x_f$ to large $x_f$
will be of interest.

In conclusion, the coherent precursor model provides a reasonable
description of the production and propagation of the object form
before the formation of the bound states.  Further tests of the model
can be carried out to confront with experiment.

The author would like to thank Dr. C. W. Wong for comments and
communications.  He also wishes to thank Drs. R. Blankenbecler,
H. W. Crater, J. C. Peng, and J. S. Wu for valuable discussions.

\end{document}